\documentclass[12pt,11lomcon,cite,epsfig,amssymb]{article}

\usepackage[cp1251]{inputenc}
\usepackage{epsfig}
\usepackage{amssymb}
\usepackage{amsmath}
\usepackage{cite}
\usepackage{datatool}
\usepackage{forloop}
\usepackage{ifthen}
\usepackage{longtable}
\usepackage{titling}

\makeatletter
\renewcommand{\section}{\@startsection{section}{1}{0pt}%
{-3.5ex plus -1ex minus -.2ex}{2.3ex plus.2ex}%
{\centering\Large\bfseries}} \makeatother

\fontsize{14}{14pt}\selectfont
\thanksmarkseries{arabic}
\begin{document}

\title{\textbf{{\large Numerical analysis of the location of spectral maxima of polarization components of synchrotron radiation in the classical theory}}}
\author{A. S. Loginov \thanks{Tomsk State University, Russia. E-mail: as\_loginov@phys.tsu.ru }, A. D. Saprykin \thanks{Tomsk State University, Russia. E-mail: alexx.saprykin.phys@gmail.com} }

\maketitle

\section*{Abstract}
Shift of higher harmonic peak in the spectrum of synchrotron radiation with increasing energy of the radiating particle is investigated by numerical method. Calculations were carried out for the  $\nu^{(max)} \leqslant 100$ for all polarization components.

\section{Introduction}

\quad \quad  At present, the theory of synchrotron radiation is a fairly well developed area of theoretical
physics. Its main elements are described in monographs (e.g. \cite{1,2,3,4,5,6,7,8}) and numerous articles

Spectral distribution of SR has one physically important feature, namely increasing energy of radiating particle leads to shift of spectral maximum to higher harmonics. This feature was predicted in works \cite{9,10} and has been confirmed by numerous experiments.

It’s convenient to represent energy $E$ of radiating particle by relativistic factor  $\gamma \,(\gamma \geqslant 1)$, or quantity   $\beta \,(0 \leqslant \beta < 1)$
\begin{equation}\label{A1}
\gamma = \frac{E}{m_0c^2} = \frac{1}{\sqrt{1 - \beta^2}}\,, \ \ \beta = \frac{v}{c}\,,
\end{equation} where $m_0$ -- the charge rest mass, $v$ is the charge orbital motion speed, $c$ -- the speed of lite.  With respect to notation of rotation frequency of radiating particle by $\omega_0$, spectrum of SR has frequencies $\omega = \omega_0 \nu$, where whole number $\nu = 1,\,2,\,3,...$ -- number of radiated harmonic.

For nonrelativistic particle which corresponds to the condition $\beta \ll 1$ (equivalently $\gamma \sim 1$) maximum of SR spectrum falls on first harmonics 	$\nu^{(max)} = 1$. With increasing energy of the particle spectral maximum shifts sequentially to 2,3,4…  harmonics and for ultra-relativistic particle (corresponding to $\gamma \gg 1$) asymptotic estimate $\nu^{(max)} \sim \gamma^3$ holds (It was shown in  \cite{9,10}). This estimate is correct for polarization component of SR.

Exact values of $\beta$ (or $\gamma$) which lead to shift of maximum in SR spectral distribution from harmonic $\nu$ to harmonic $\nu + 1$ can be carried out by numerical calculations. within the framework of classical theory such calculations have been carried out in the work \cite{11} for $\nu \leqslant 15$ for the total radiated power and the linear polarization component of SR.
This paper presents the results of numerical analysis for $\nu \leqslant 100$ and all polarization component of SR.

\section{Spectral distribution of SR polarization component in upper half-space}

\quad \quad To put a numerical analysis of the problem considered in this article we present here some well-known expression of the physical characteristics SR in the classical theory which can be found in \cite{1,2,3,4,5,6,7,8}.

The spectral - angular distribution of the radiation power of the SR polarization component can be represented by
\begin{equation}\label{B1}
W_s = W\sum_{\nu = 1}^\infty \int_0^\pi f_s (\beta;\, \nu,\,\theta) \sin\theta d \theta.
\end{equation} Here, the following notation is used: $\theta$ is the angle between the control magnetic field strength and the radiation field pulse,  $W$ -- the total radiated power of unpolarized radiation, which can be written as
\begin{equation}\label{B2}
W = \frac{2}{3} \frac{ce^{2}}{R^{2}} (\gamma^2 - 1)^2 =
\frac{2 e^{4}H^{2}(\gamma^2 - 1)}{3 m_{0}^{2}c^{3}} \, ,
\end{equation} where $e$ -- the particle charge;; $R$ -- orbit radius;; $H$ -- is the control field strength. The index $s$ numbers the polarization components:: $s = 2$ corresponds to  $\sigma$ - component of linear polarization;  $s = 3$ corresponds to  $\pi$ - component of linear polarization; $s = 1$ corresponds to right-hand circular polarization; $s = - 1$ -- corresponds to left-hand circular polarization;; $s = 0$ corresponds to the power of unpolarized radiation. The functions $f_s (\beta;\, \nu,\,\theta)$ have the form
\begin{equation}\notag
f_2 (\beta;\, \nu,\,\theta) = \frac{3 \nu^2}{2 \gamma^4} J _{\nu}'\,^2(x) ; \ \ f_3 (\beta;\, \nu,\,\theta) = \frac{3 \nu^2}{2 \gamma^4} \frac{\cos^2 \theta }{\beta^2 \sin^2 \theta} J^2_{\nu}(x) ;
\end{equation}
\begin{equation}\notag
f_{\pm 1} (\beta;\, \nu,\,\theta) = \frac{3 \nu^2}{4 \gamma^4} \left[J _{\nu}'(x) \pm \varepsilon\frac{\cos \theta }{\beta \sin \theta} J_{\nu}(x) \right]^2 ; \ \ x = \nu \beta\sin\theta ; \ \ \varepsilon = - \frac{e}{|e|} ;
\end{equation}
\begin{equation}\label{B3}
f_{0} (\beta;\, \nu,\,\theta) = f_{2} (\beta;\, \nu,\,\theta) + f_{3} (\beta;\, \nu,\,\theta) = f_{1} (\beta;\, \nu,\,\theta) + f_{- 1} (\beta;\, \nu,\,\theta)\,.
\end{equation} Here, $J_{\nu}(x)$ and $J _{\nu}'(x)$  are the Bessel functions and their derivatives. In what follows, the case of an electron is considered, which corresponds to $\varepsilon = 1$.

It is well known \cite{1,2,3,4,5,6,7} that the angle range $0\leqslant \theta < \pi/2$ (this range will be called the upper half-space) is dominated by right-hand circular polarization, and the angle range $\pi/2 < \theta \leqslant\pi$ (this range will be called the lower half-space) is dominated by left-hand circular polarization (exact quantitative characteristics of SR properties were first obtained in \cite{12,13,14,15}). However, if we integrate in (\ref{B1})) over $\theta \ \ (0\leqslant \theta \leqslant \pi)$ then the differences in the spectral distribution of right- and left-hand circular polarizations disappear. To reveal these differences, a new form of expression (\ref{B1})) has been proposed in \cite{16} It can be represented as
\begin{equation}\notag
W_s = W_s^{(+)} + W_s^{(-)}\,, \ \  W_s^{(\pm)} = W \sum_{\nu = 1}^\infty F_s^{(\pm)}(\beta;\,\nu) \,;
\end{equation}
\begin{equation}\label{B4}
F_s^{(+)}(\beta;\,\nu) = \int_0^{\pi/2} f_s (\beta;\, \nu,\,\theta) \sin\theta d \theta\,, \ \ F_s^{(-)}(\beta;\,\nu) = \int_{\pi/2}^{\pi} f_s (\beta;\, \nu,\,\theta) \sin\theta d \theta\,,
\end{equation} and it suffices to study the properties of functions $F_s^{(+)}(\beta;\,\nu)$ due to the evident relations
\begin{equation}\notag
F_s^{(-)}(\beta;\,\nu) = F_s^{(+)}(\beta;\,\nu)\,, \ \ s = 0,\,2,\,3;
\end{equation}
\begin{equation}\label{B5}
F_1^{(-)}(\beta;\,\nu) = F_{- 1}^{(+)}(\beta;\,\nu)\,, \ \ F_{-1}^{(-)}(\beta;\,\nu) = F_{ 1}^{(+)}(\beta;\,\nu)\,.
\end{equation}

Integration over $\theta$ in the upper half-space $0\leqslant \theta \leqslant \pi/2$ in (\ref{B4}) can be carried out exactly, yielding the expressions
\begin{equation}\notag
F_2^{(+)}(\beta;\,\nu) = \frac{3 \nu}{4\,\gamma\,^4 \beta^3} \left[2 \beta^2 J_{2 \nu}'(2 \nu \beta) + \beta^2
\int_{0}^{2 \nu \beta} J_{2 \nu} (x) d x - 2 \nu \beta \int_{0}^{2 \nu \beta} \frac{J_{2 \nu} (x)}{x} d x \right]\,,
\end{equation}
\begin{equation}\notag
F_3^{(+)}(\beta;\,\nu) = \frac{3 \nu}{4\,\gamma\,^4 \beta^3}\left[2 \nu \beta \int_{0}^{2 \nu \beta} \frac{J_{2
\nu} (x)}{x} d x - \int_{0}^{2 \nu \beta} J_{2 \nu} (x) d x \right]\,,
\end{equation}
\begin{equation}\notag
F_0^{(+)}(\beta;\,\nu) = F_2^{(+)}(\beta;\,\nu) + F_3^{(+)}(\beta;\,\nu) = \frac{3 \nu}{4\,\gamma\,^4 \beta^3} \left[2 \beta^2 J_{2 \nu}'(2 \nu \beta) - (1 - \beta^2) \int_{0}^{2 \nu \beta} J_{2 \nu} (x) d x \right]\,,
\end{equation}
\begin{equation}\label{B6}
F_{\pm 1}^{(+)}(\beta;\,\nu) = \frac{1}{2} F_0^{(+)}(\beta;\,\nu) \pm  \frac{3 \nu J\,^2_{\nu}(\nu \beta)}{4\,\gamma\,^4 \beta^2}\,.
\end{equation}

\section{Numerical analysis of the shift of the maximum in the SR spectrum with increasing the energy of the radiating particle}

\quad \quad Expressions (\ ref {B6}) make possible to translate into formal language the task of analyzing the shift of SR spectral maximum at change of radiating particle energy $E$.

As is known, in the non-relativistic limit (which corresponds to $ \beta \rightarrow 0 $ and $ \gamma \rightarrow 1 $) maxima of any components $W_s^{(+)}$ in the spectral distribution occur in first harmonic ($\nu^{(max)}_s = 1$). With increasing the energy of particle (increase of $ \beta $ or, respectively, $ \gamma $) maximum in the spectrum of $ W_s ^ {(+)} $ goes to the second harmonic.

Obviously, the value of $ \beta = \beta ^ {(s)} _ 1$, which is carried out at the maximum shift in the spectrum of $ W_s ^ {(+)} $ to the second harmonic is the root of the equation
\begin{equation}\label{C1}
F_s^{(+)}(\beta;\,1) = F_s^{(+)}(\beta;\,2)\,.
\end{equation}
So for $ 0 \leqslant \beta <\beta ^ {(s)} _ 1$ (or $ 1 \leqslant \gamma <\gamma ^ {(s)} _ 1$) the maximum in spectrum of $ W_s ^ {( +)} $ falls on the first harmonic

With increasing $\beta$ (or $\gamma$) maximum in spectrum of $W_s^{(+)}$ shift to second harmonic. The value $\beta = \beta^{(s)}_2$, corresponding to this shift, is root of equation
\begin{equation}\label{C2}
F_s^{(+)}(\beta;\,2) = F_s^{(+)}(\beta;\,3) \,.
\end{equation} So for $\beta^{(s)}_1 \leqslant \beta < \beta^{(s)}_2$  (or $\gamma^{(s)}_1 \leqslant \gamma < \gamma^{(s)}_2$) maximum in spectrum of $W_s^{(+)}$ falls on second harmonic.

In general, for a fixed $ s $ determining the sequence of numbers $ \beta = \beta ^ {(s)} _\nu $ as the roots of equations
\begin{equation}\label{C3}
F_s^{(+)}(\beta;\,\nu) = F_s^{(+)}(\beta;\,\nu + 1) \,,
\end{equation} It can be found that in energy range
\begin{equation}\label{C4}
\beta^{(s)}_{\nu - 1} \leqslant \beta < \beta^{(s)}_{\nu} \ \ \left(\mbox{respectively} \,\, \gamma^{(s)}_{\nu - 1} \leqslant \gamma < \gamma^{(s)}_\nu \right)
\end{equation} maximum in spectrum of $W_s^{(+)}$ falls on harmonic $\nu$. 

The following Table for $ \nu \leqslant $ 100 for each polarization $ s $  $\beta ^ {(s)} _ {\nu} $ and $ \gamma ^ {(s)} _ {\nu} $ are given. It should be noted that following inequalities are correct
\begin{equation}\label{C5}
\gamma^{(3)}_{\nu} > \gamma^{(1)}_{\nu} > \gamma^{(0)}_{\nu} > \gamma^{(2)}_{\nu} > \gamma^{(- 1)}_{\nu} \ \  \left(\mbox{respectively} \,\, \beta^{(3)}_{\nu} > \beta^{(1)}_{\nu} > \beta^{(0)}_{\nu} > \beta^{(2)}_{\nu} > \beta^{(- 1)}_{\nu} \right).
\end{equation} It is also easy to see from Table, that since $\nu^{(max)}_{- 1} (\beta) = 8$ inequalities, obtained in [16] for ultra-relativistic particle, hold.

In the ultra-relativistic case ($ 1 \ll \gamma $) for $ \nu ^ {(max)} _ s = 1 $ asymptotic formulas are known.
\begin{equation}\label{C6}
\nu^{(max)}_s \approx a_s \gamma^3 \,,
\end{equation} where $a_s$ -- numbers, approximate values of ones are equal (see [16]) to
\begin{equation}\label{C7}
a_3 \approx 0,215881 \,, \ \ a_1 \approx 0,372875 \,, \ \ a_0 \approx 0,428718 \,, \ \ a_2 \approx 0,503287 \,, \ \ a_{- 1} \approx 0,783608 \,.
\end{equation}

If $ \nu ^ {(max)} _ s = $ 100 value corresponding $ \gamma $ determined by the asymptotic formulas (\ref {C6}), they may differ from the table of values in no more than fourth place.

\DTLsettabseparator
\begin{filecontents*}{betas_max.txt}
1	0,71493	1,43020	0,88148	2,11768	0,76354	1,54856	0,58762	1,23588	0,78799	1,62420	
2	0,80568	1,68819	0,91747	2,51376	0,83681	1,82649	0,72085	1,44281	0,85442	1,92461	
3	0,84815	1,88768	0,93305	2,77982	0,87113	2,03643	0,78352	1,60936	0,88498	2,14761	
4	0,87340	2,05334	0,94249	2,99182	0,89174	2,20971	0,82060	1,74978	0,90325	2,33037	
5	0,89036	2,19655	0,94902	3,17254	0,90572	2,35920	0,84538	1,87209	0,91563	2,48749	
6	0,90265	2,32359	0,95390	3,33211	0,91594	2,49175	0,86325	1,98108	0,92469	2,62652	
7	0,91203	2,43835	0,95773	3,47611	0,92378	2,61147	0,87682	2,07979	0,93164	2,75195	
8	0,91946	2,54340	0,96082	3,60802	0,93002	2,72110	0,88753	2,17033	0,93719	2,86671	
9	0,92552	2,64056	0,96340	3,73018	0,93513	2,82252	0,89621	2,25418	0,94173	2,97282	
10	0,93056	2,73115	0,96557	3,84428	0,93941	2,91712	0,90343	2,33243	0,94553	3,07177	
11	0,93483	2,81619	0,96745	3,95154	0,94304	3,00594	0,90953	2,40593	0,94876	3,16465	
12	0,93851	2,89644	0,96908	4,05295	0,94618	3,08979	0,91477	2,47533	0,95156	3,25232	
13	0,94171	2,97253	0,97052	4,14924	0,94892	3,16932	0,91932	2,54115	0,95400	3,33546	
14	0,94453	3,04495	0,97180	4,24104	0,95133	3,24504	0,92331	2,60383	0,95616	3,41461	
15	0,94704	3,11411	0,97295	4,32884	0,95348	3,31738	0,92686	2,66370	0,95808	3,49022	
16	0,94928	3,18037	0,97399	4,41306	0,95541	3,38669	0,93002	2,72106	0,95980	3,56266	
17	0,95130	3,24400	0,97493	4,49404	0,95715	3,45327	0,93287	2,77615	0,96136	3,63224	
18	0,95313	3,30525	0,97579	4,57209	0,95873	3,51737	0,93545	2,82919	0,96277	3,69924	
19	0,95480	3,36433	0,97658	4,64745	0,96018	3,57922	0,93780	2,88035	0,96406	3,76388	
20	0,95633	3,42142	0,97730	4,72036	0,96150	3,63899	0,93995	2,92980	0,96525	3,82636	
21	0,95774	3,47668	0,97797	4,79099	0,96272	3,69687	0,94192	2,97766	0,96634	3,88684	
22	0,95904	3,53025	0,97860	4,85954	0,96385	3,75298	0,94374	3,02407	0,96735	3,94549	
23	0,96025	3,58226	0,97918	4,92614	0,96489	3,80746	0,94543	3,06911	0,96829	4,00244	
24	0,96137	3,63281	0,97972	4,99092	0,96587	3,86043	0,94700	3,11290	0,96916	4,05780	
25	0,96241	3,68200	0,98023	5,05401	0,96678	3,91198	0,94846	3,15550	0,96997	4,11168	
26	0,96339	3,72992	0,98071	5,11552	0,96763	3,96220	0,94982	3,19701	0,97074	4,16417	
27	0,96431	3,77664	0,98116	5,17554	0,96843	4,01118	0,95110	3,23748	0,97145	4,21537	
28	0,96517	3,82225	0,98158	5,23416	0,96918	4,05899	0,95230	3,27699	0,97213	4,26535	
29	0,96598	3,86680	0,98198	5,29146	0,96989	4,10570	0,95343	3,31557	0,97277	4,31417	
30	0,96675	3,91035	0,98236	5,34750	0,97055	4,15137	0,95450	3,35330	0,97337	4,36191	
31	0,96747	3,95296	0,98272	5,40237	0,97119	4,19606	0,95551	3,39020	0,97394	4,40862	
32	0,96816	3,99467	0,98306	5,45611	0,97179	4,23981	0,95646	3,42633	0,97447	4,45436	
33	0,96881	4,03554	0,98339	5,50879	0,97236	4,28268	0,95737	3,46172	0,97499	4,49918	
34	0,96943	4,07560	0,98370	5,56045	0,97290	4,32470	0,95823	3,49642	0,97547	4,54311	
35	0,97002	4,11489	0,98399	5,61115	0,97342	4,36593	0,95905	3,53045	0,97594	4,58620	
36	0,97058	4,15345	0,98427	5,66093	0,97391	4,40639	0,95983	3,56384	0,97638	4,62850	
37	0,97112	4,19131	0,98454	5,70982	0,97438	4,44611	0,96057	3,59663	0,97681	4,67004	
38	0,97163	4,22850	0,98480	5,75787	0,97483	4,48514	0,96128	3,62884	0,97721	4,71084	
39	0,97213	4,26505	0,98505	5,80511	0,97526	4,52350	0,96196	3,66049	0,97760	4,75095	
40	0,97260	4,30099	0,98529	5,85158	0,97567	4,56123	0,96261	3,69161	0,97797	4,79039	
41	0,97305	4,33634	0,98552	5,89731	0,97607	4,59833	0,96324	3,72222	0,97833	4,82919	
42	0,97348	4,37113	0,98574	5,94232	0,97645	4,63485	0,96384	3,75234	0,97867	4,86737	
43	0,97390	4,40537	0,98595	5,98665	0,97681	4,67080	0,96441	3,78200	0,97900	4,90496	
44	0,97430	4,43909	0,98616	6,03032	0,97716	4,70620	0,96496	3,81120	0,97931	4,94198	
45	0,97468	4,47231	0,98635	6,07335	0,97750	4,74108	0,96550	3,83996	0,97962	4,97845	
46	0,97505	4,50505	0,98654	6,11576	0,97783	4,77546	0,96601	3,86830	0,97991	5,01439	
47	0,97541	4,53731	0,98673	6,15759	0,97814	4,80934	0,96650	3,89624	0,98020	5,04982	
48	0,97576	4,56913	0,98690	6,19884	0,97845	4,84276	0,96698	3,92379	0,98047	5,08476	
49	0,97609	4,60051	0,98707	6,23954	0,97874	4,87571	0,96744	3,95096	0,98074	5,11923	
50	0,97641	4,63148	0,98724	6,27970	0,97903	4,90823	0,96788	3,97776	0,98099	5,15323	
51	0,97672	4,66203	0,98740	6,31935	0,97930	4,94033	0,96831	4,00422	0,98124	5,18679	
52	0,97703	4,69219	0,98756	6,35850	0,97957	4,97201	0,96873	4,03033	0,98148	5,21992	
53	0,97732	4,72197	0,98771	6,39716	0,97982	5,00329	0,96913	4,05611	0,98171	5,25264	
54	0,97760	4,75138	0,98785	6,43535	0,98007	5,03418	0,96952	4,08157	0,98194	5,28495	
55	0,97788	4,78043	0,98800	6,47309	0,98031	5,06470	0,96990	4,10672	0,98215	5,31686	
56	0,97814	4,80914	0,98813	6,51038	0,98055	5,09486	0,97027	4,13157	0,98237	5,34840	
57	0,97840	4,83751	0,98827	6,54724	0,98078	5,12466	0,97062	4,15612	0,98257	5,37957	
58	0,97865	4,86555	0,98840	6,58369	0,98100	5,15413	0,97097	4,18039	0,98277	5,41038	
59	0,97890	4,89327	0,98852	6,61972	0,98121	5,18325	0,97130	4,20439	0,98297	5,44085	
60	0,97913	4,92068	0,98865	6,65537	0,98142	5,21206	0,97163	4,22811	0,98315	5,47097	
61	0,97936	4,94779	0,98877	6,69063	0,98163	5,24055	0,97195	4,25158	0,98334	5,50077	
62	0,97959	4,97461	0,98888	6,72551	0,98182	5,26873	0,97225	4,27479	0,98352	5,53024	
63	0,97981	5,00114	0,98900	6,76004	0,98202	5,29661	0,97255	4,29776	0,98369	5,55941	
64	0,98002	5,02740	0,98911	6,79420	0,98220	5,32420	0,97285	4,32048	0,98386	5,58827	
65	0,98023	5,05338	0,98922	6,82802	0,98239	5,35151	0,97313	4,34297	0,98402	5,61683	
66	0,98043	5,07910	0,98932	6,86151	0,98256	5,37855	0,97341	4,36523	0,98419	5,64511	
67	0,98062	5,10457	0,98943	6,89466	0,98274	5,40531	0,97368	4,38727	0,98434	5,67310	
68	0,98082	5,12978	0,98953	6,92750	0,98291	5,43181	0,97394	4,40909	0,98450	5,70082	
69	0,98100	5,15475	0,98963	6,96002	0,98307	5,45806	0,97420	4,43070	0,98464	5,72827	
70	0,98119	5,17948	0,98972	6,99223	0,98323	5,48405	0,97445	4,45210	0,98479	5,75546	
71	0,98136	5,20398	0,98981	7,02415	0,98339	5,50980	0,97469	4,47330	0,98493	5,78240	
72	0,98154	5,22825	0,98991	7,05577	0,98355	5,53531	0,97493	4,49430	0,98507	5,80908	
73	0,98171	5,25229	0,99000	7,08711	0,98370	5,56059	0,97517	4,51511	0,98521	5,83553	
74	0,98187	5,27612	0,99008	7,11817	0,98384	5,58564	0,97539	4,53573	0,98534	5,86173	
75	0,98204	5,29974	0,99017	7,14896	0,98399	5,61047	0,97562	4,55616	0,98547	5,88770	
76	0,98220	5,32315	0,99025	7,17948	0,98413	5,63508	0,97583	4,57642	0,98560	5,91345	
77	0,98235	5,34635	0,99033	7,20974	0,98427	5,65948	0,97605	4,59650	0,98572	5,93897	
78	0,98250	5,36936	0,99042	7,23975	0,98440	5,68367	0,97626	4,61640	0,98584	5,96427	
79	0,98265	5,39217	0,99049	7,26950	0,98453	5,70765	0,97646	4,63614	0,98596	5,98936	
80	0,98280	5,41479	0,99057	7,29901	0,98466	5,73144	0,97666	4,65571	0,98608	6,01424	
81	0,98294	5,43722	0,99065	7,32828	0,98479	5,75502	0,97686	4,67512	0,98619	6,03892	
82	0,98308	5,45947	0,99072	7,35732	0,98491	5,77842	0,97705	4,69437	0,98631	6,06339	
83	0,98322	5,48154	0,99079	7,38612	0,98503	5,80163	0,97724	4,71347	0,98642	6,08767	
84	0,98335	5,50344	0,99086	7,41470	0,98515	5,82465	0,97742	4,73241	0,98652	6,11176	
85	0,98349	5,52516	0,99093	7,44305	0,98527	5,84750	0,97760	4,75121	0,98663	6,13565	
86	0,98361	5,54672	0,99100	7,47119	0,98538	5,87016	0,97778	4,76986	0,98673	6,15937	
87	0,98374	5,56810	0,99107	7,49912	0,98550	5,89265	0,97795	4,78836	0,98683	6,18290	
88	0,98387	5,58933	0,99114	7,52683	0,98561	5,91497	0,97812	4,80672	0,98693	6,20625	
89	0,98399	5,61039	0,99120	7,55435	0,98571	5,93713	0,97829	4,82495	0,98703	6,22943	
90	0,98411	5,63130	0,99126	7,58165	0,98582	5,95912	0,97845	4,84303	0,98713	6,25243	
91	0,98422	5,65206	0,99133	7,60877	0,98592	5,98095	0,97861	4,86099	0,98722	6,27527	
92	0,98434	5,67266	0,99139	7,63568	0,98603	6,00262	0,97877	4,87882	0,98731	6,29794	
93	0,98445	5,69312	0,99145	7,66241	0,98613	6,02413	0,97892	4,89651	0,98740	6,32045	
94	0,98456	5,71343	0,99151	7,68895	0,98622	6,04550	0,97908	4,91408	0,98758	6,36500	
95	0,98467	5,73360	0,99157	7,71530	0,98632	6,06671	0,97923	4,93153	0,98758	6,36500	
96	0,98478	5,75363	0,99162	7,74148	0,98642	6,08777	0,97937	4,94885	0,98775	6,40892	
97	0,98489	5,77351	0,99168	7,76747	0,98651	6,10869	0,97952	4,96606	0,98775	6,40892	
98	0,98499	5,79327	0,99173	7,79329	0,98660	6,12947	0,97966	4,98314	0,98784	6,43066	
99	0,98509	5,81289	0,99179	7,81894	0,98669	6,15011	0,97980	5,00012	0,98792	6,45225	
100	0,98519	5,83238	0,99184	7,84442	0,98678	6,17061	0,97993	5,01697	0,98800	6,47370	
											
											
											
\end{filecontents*}
\DTLloaddb[noheader,keys = {a,b,c,d,e,f,g,h,i,j,k},omitlines = 0]{data}{betas_max.txt}
\setcounter{LTchunksize}{40}
\begin{longtable}[c]{|*{11}{c|}}
\caption{ Boundary values of energy (corresponding to $\beta^{(s)}_{\nu}$) for spectral maximum shift }\\ \hline
 $\nu$ & $\beta^{(2)}_{\nu}$ & $\gamma^{(2)}_{\nu}$ & $\beta^{(3)}_{\nu}$ & $\gamma^{(3)}_{\nu}$ & $\beta^{(0)}_{\nu}$ & $\gamma^{(0)}_{\nu}$&$\beta^{(-1)}_{\nu}$ & $\gamma^{(-1)}_{\nu}$&$\beta^{(1)}_{\nu}$ & $\gamma^{(1)}_{\nu}$\\ \hline \endhead   \hline \endfoot
\DTLforeach[\value{DTLrowi}<101]{data}{ \a =a,\b=b,\c=c,\d=d,\e=e,\f=f,\g=g,\h=h,\i=i,\j=j,\k=k}{\DTLiffirstrow{}{\\ } \a & \b & \c & \d & \e &\f & \g & \h & \i & \j & \k  }
\end{longtable}

 \DTLdeletedb{data}
 \section*{Acknowledgements}
  The work of Loginov and Saprykin is  supported by Tomsk State University Competitiveness Improvement Program.

\end{document}